\documentstyle[preprint,aps]{revtex}
\tighten
\draft
\widetext
\preprint{YITP-SB-01-40, NYU-TH/01/07/10}
\bigskip
\bigskip

\newcommand{\EQ}{\begin{equation}}
\newcommand{\EN}{\end{equation}}
\newcommand{\bea}{\begin{eqnarray}}
\newcommand{\ena}{\end{eqnarray}}
\newcommand{\eea}{\end{eqnarray}}

\newcommand{\bi}{\begin{itemize}}
\newcommand{\ei}{\end{itemize}}

\begin{document}
\title{Gravity on a 3-brane in 6D Bulk}
\medskip
\author{
Olindo Corradini$^1$\footnote{E-mail: olindo@insti.physics.sunysb.edu},
Alberto Iglesias$^1$\footnote{E-mail: iglesias@insti.physics.sunysb.edu},
Zurab Kakushadze$^{1,2}$\footnote{E-mail: zurab@insti.physics.sunysb.edu}
and Peter Langfelder$^1$\footnote{E-mail: plangfel@insti.physics.sunysb.edu}}
\bigskip
\address{$^1$C.N. Yang Institute for Theoretical Physics\\
State University of New York, Stony Brook, NY 11794\\
$^2$Department of Physics, New York University, New York, NY 10003}

\date{August 8, 2001}
\bigskip
\medskip
\maketitle

\begin{abstract}
{}We study gravity in codimension-2 brane world scenarios with infinite
volume extra dimensions. In particular, we consider the case where the brane
has non-zero tension. The extra space then is a two-dimensional ``wedge''
with a deficit angle. In such backgrounds we can effectively
have the Einstein-Hilbert
term on the brane at the classical level if we include higher curvature
(Gauss-Bonnet) terms in the bulk. Alternatively, such a term would be 
generated at the quantum level if the brane matter is not conformal. 
We study (linearized) gravity in the presence of the
Einstein-Hilbert term on the brane in such backgrounds. 
We find that, just as in the original codimension-2 
Dvali-Gabadadze model with a tensionless brane, 
gravity is almost completely localized on the brane with
ultra-light modes penetrating into the bulk.
\end{abstract}

\section{Introduction and Summary}

{}In the Brane World scenario the Standard Model gauge and matter fields
are assumed to be localized on
branes (or an intersection thereof), while gravity lives in a larger
dimensional bulk of space-time
\cite{early,BK,polchi,witt,lyk,shif,TeV,dienes,3gen,anto,ST,BW,Gog,RS,kogan,DGP,DG,alberto}.
There is a big difference between the footings on which
gauge plus matter fields and gravity come in this picture\footnote{This, at
least in some sense, might not be an unwelcome feature - see, {\em e.g.},
\cite{witt,TeV,BW}.}. Thus, for instance, if gauge and matter fields are
localized on D-branes \cite{polchi}, they propagate only in the directions
along the D-brane world-volume. Gravity, however, is generically
not confined to the branes - even if we have a graviton zero mode localized
on the brane as in \cite{RS}, where the volume of the extra dimension is
finite, massive graviton modes are still free to propagate in the bulk.

{}On the other hand, as was originally proposed in \cite{DG},
in the cases with infinite volume extra dimensions 
\cite{GRS,CEH,DGP0,witten,DVALI,zura}, we can
have almost completely localized gravity on higher codimension
($\delta$-function-like) branes with the ultra-light modes penetrating into the
bulk\footnote{A rather different mechanism was also proposed in \cite{alberto},
which leads to a complete localization of gravity on a codimension-1 brane
with no (perturbative) modes propagating in the bulk.}. As was explained in
\cite{DG}, this dramatic modification of gravity
in higher codimension models with infinite volume extra dimensions
is due to the 
Einstein-Hilbert term on the brane, which, as was originally pointed out in
\cite{DGP,DG}, is induced via loops of
non-conformal brane matter. 

{}In the original models of 
\cite{DG} the brane is tensionless,
so that the $D$-dimensional space-time is Minkowski. The purpose of this paper
is to consider such models with non-zero tension brane. In this case the
bulk is no longer flat (but the brane is). In fact, at the origin 
of the extra space (that is, at the location of the brane) we have
curvature singularities 
in these models. In codimension-3 and higher cases these
curvature singularities are difficult to deal with. However, in the 
codimension-2 case, which we focus on in this paper, the singularity is
$\delta$-function-like. That is, the space away from the brane is locally flat,
and all the curvature is concentrated at the location of the 
brane. In fact, the extra space
in this case is a two-dimensional ``wedge'' with a deficit angle, which 
depends on the brane tension.

{}Thus, in this paper we analyze brane world gravity in such codimension-2 
backgrounds\footnote{Certain codimension-2 solutions were discussed in
\cite{CP,GS,olzu2}.}. The Einstein-Hilbert term on the brane can 
effectively be present 
classically if we include higher curvature (Gauss-Bonnet) terms in the bulk.
Alternatively, such a term on the brane is generated at the quantum level
if the brane matter is not conformal \cite{DGP,DG}. 
We study gravity in the presence of the
Einstein-Hilbert term on the brane in such backgrounds. 
We find that, just as in the original codimension-2 
Dvali-Gabadadze model with a tensionless brane \cite{DG}, 
we still have almost complete localization of gravity
on the brane. Thus, in the case of a non-zero tension
3-brane in infinite volume 6-dimensional space we have 4-dimensional gravity
on the brane with ultra-light modes penetrating into the bulk.

{}The remainder of this paper is organized as follow. In section II we present
the model along with the aforementioned background solution. In section III we
study small fluctuations around the solution in the presence of brane matter
sources.

\section{The Model}

{}In this section we discuss a brane world model with a codimension-2 brane 
embedded in a $D$-dimensional bulk space. (For calculational convenience
we will keep the number of space-time dimensions $D$ unspecified, but we are
mostly interested in the case $D=6$, where the brane is a 3-brane.) The
action for this model is given by:
\EQ
 S=-f\int_{\Sigma}\ d^{D-2}x\sqrt{-\widehat G}+M_P^{D-2}\int\ d^D x\sqrt{-G}
 \biggl[R+\lambda\biggl(R^2-4R_{MN}^2+R^2_{MNRS}\biggr)\biggr]~.
 \label{model}
\EN
Here $M_P$ is the (reduced) $D-$dimensional Planck mass; $\Sigma$ is a 
$\delta$-function-like codimension-2 source brane, which is 
a hypersurface $x^i=0$ ($x^i$, $i=1,2$, are the two spatial coordinates 
transverse to the brane); the tension $f$ of the brane is assumed to be
positive; also, 
\begin{equation}
 {\widehat G}_{\mu\nu}\equiv
 {\delta_\mu}^M{\delta_\nu}^N G_{MN}\Big|_\Sigma~,
\end{equation} 
where $x^\mu$ are the $(D-2)$ coordinates along the brane (the $D$-dimensional
coordinates are given by $x^M=(x^\mu,x^i)$, and the signature of the 
$D$-dimensional metric is $(-,+,\dots,+)$); finally, the higher curvature
terms in the bulk action are chosen in the Gauss-Bonnet combination, and
the Gauss-Bonnet coupling $\lambda$ is {\em a priori} a free parameter
(which, as we will see below, is restricted to be non-negative by
unitarity considerations).

{}The equations of motion following from the action (\ref{model}) are given by:
\begin{eqnarray}
 &&R_{MN}-\frac{1}{2}\ G_{MN}\biggl[R+\lambda\biggl(R^2-4R_{MN}^2
 +R^2_{MNRS}\biggr)
 \biggr]+\nonumber\\
 &&+2\lambda\biggl(R R_{MN}-2R_{MS} R^S{}_N+R_{MRST} R_N{}^{RST}
 -2R^{RS} R_{MRNS}\biggr)+\nonumber\\
 &&+ \displaystyle{
 {1\over 2}{\sqrt{-{\widehat G}}\over\sqrt{-G}}}
 {\delta_M}^\mu {\delta_N}^\nu 
 {\widehat G}_{\mu\nu}{\widetilde f} \delta^{(2)}(x^i)=0~,
 \label{EoM}
\end{eqnarray}
where ${\widetilde f}\equiv f/M_P^{D-2}$.

{}Consider the following ans{\"a}tz for the 
metric\footnote{A similar solution was recently discussed
in \cite{olzu2}.}
\EQ
 ds^2=\eta_{\mu\nu}dx^\mu dx^\nu+\exp(2\omega)~\delta_{ij}dx^i dx^j~,
\label{ansatz}
\EN
where $\omega$ is a function of $x^i$ but is independent of $x^\mu$. With this
ans{\"a}tz we have:
\begin{eqnarray}
 &&R_{\mu\nu}=R_{\mu i}=0~,~~~
 R_{ij}={\widetilde R}_{ij}={1\over 2} {\widetilde G}_{ij}
 {\widetilde R}~,\\
 &&\sqrt{\widetilde G}{\widetilde R}={\widetilde f}~\delta^{(2)}(x^i)~,
\end{eqnarray}
where ${\widetilde R}$ and ${\widetilde R}_{ij}$ are the 2-dimensional
Ricci scalar respectively Ricci tensor constructed from the 2-dimensional
metric
\begin{equation}
 {\widetilde G}_{ij}=\exp(2\omega)~\delta_{ij}~.
\end{equation}
Since this metric is conformally flat, we have 
$\sqrt{\widetilde G}{\widetilde R}=-2\partial^i\partial_i \omega$ (where the
indices are lowered and raised with $\delta_{ij}$ and $\delta^{ij}$,
respectively), so we have:
\EQ
 \partial^i\partial_i \omega =-{1\over 2}{\widetilde f} \delta^{(2)}(x^i)~.
\label{EoM:extra}
\EN
The solution to this equation is given by:
\EQ
 \omega(x^i)=-{1\over 8\pi}{\widetilde f}~{\rm ln}
 \biggl({x^2\over a^2}\biggr)~,
\label{EoM:solution}
\EN
where $x^2\equiv x^i x_i$, and $a$ is an integration constant.

{}Let us go to the polar coordinates $(\rho,\phi)$: $x^1=\rho\cos(\phi)$,
$x^2=\rho\sin(\phi)$ ($\rho$ takes values from 0 to $\infty$, while $\phi$
takes values from 0 to $2\pi$). In these coordinates the two dimensional
metric is given by
\begin{equation}
 d{\widetilde s}_2^2=\left({a^2\over \rho^2}\right)^\nu\left[(d\rho)^2+
 \rho^2(d\phi)^2\right]~,
\end{equation}
where
\begin{equation}
 \nu\equiv{1\over 4\pi}{\widetilde f}~.
\end{equation}
Let us change the coordinates to $(r,\phi)$, where 
\begin{equation}
 r\equiv{1\over {1-\nu}}~a^\nu\rho^{1-\nu}~,
\end{equation}
where we are assuming that $\nu<1$. Then we have
\begin{equation}
 d{\widetilde s}_2^2=(dr)^2+
 \exp(-2\beta)~r^2(d\phi)^2~,
\end{equation}
where
\begin{equation}
 \exp(-\beta)\equiv 1-\nu~.
\end{equation}
Thus, we see that the $D$-dimensional space-time in this solution is
the $(D-2)$-dimensional Minkowski space times a 2-dimensional ``wedge''
with the deficit angle
\begin{equation}
 \theta=2\pi[1-\exp(-\beta)]={{\widetilde f}\over 2}~.
\end{equation}
That is, the brane is flat for a continuous range
of values of the brane tension $f$. Note that for the critical
value $f_c$ of the brane tension, where
\EQ
 f_c\equiv 4\pi M_P^{D-2}~,
 \label{tension}
\EN
the deficit angle is $2\pi$. Thus, we have a flat solution
for the brane tension $0<f<f_c$. Note that the Gauss-Bonnet coupling does
not enter in this solution due to the fact that the space-time is
factorizable, the curvature comes from the 2-dimensional wedge (in fact,
the origin thereof), and the Gauss-Bonnet combination is trivial in
two dimensions. 
However, as we will see in the following, the higher curvature
bulk terms in this model do contribute to fluctuations around the background,
and, in fact, effectively give rise to the $(D-2)$-dimensional 
Einstein-Hilbert term on the brane.

\section{Brane World Gravity}

{}In this section we study gravity in the brane world solution
discussed in the previous section. Thus, let us consider small fluctuations
around the solution
\EQ
 G_{MN}=G^{(0)}_{MN}+h_{MN}
 \label{metric}
\EN
where $G^{(0)}_{MN}$ is the background metric
\bea
 G_{MN}^{(0)}=\left[
 \begin{array}{cc}
 \eta_{\mu\nu} & 0 \\
 0 & \exp(2\omega)\delta_{ij}\\
 \end{array}
 \right]~.
 \label{background}
\eea
The $(D-2)$-dimensional graviton $H_{\mu\nu}\equiv h_{\mu\nu}$
couples to the matter localized on the brane via
\bea
 S_{\rm int}={1\over 2}\int_{\Sigma}d^{D-2}x\ T_{\mu\nu}H^{\mu\nu}~,
 \label{interaction}
\eea
where $T_{\mu\nu}$ is the conserved energy-momentum tensor for the
matter localized on the brane:
\bea 
 \partial^\mu T_{\mu\nu}=0~.
\eea
The equations of motion read
\begin{eqnarray}
 &&R_{MN}-\frac{1}{2}\ G_{MN}\biggl[R+\lambda\biggl(R^2-4R_{MN}^2
 +R^2_{MNRS}\biggr)
 \biggr]+\nonumber\\
 &&+2\lambda\biggl(R R_{MN}-2R_{MS} R^S{}_N+R_{MRST} R_N{}^{RST}
 -2R^{RS} R_{MRNS}\biggr)+\nonumber\\
 &&+ \displaystyle{
 {1\over2\sqrt{-G}}}
 {\delta_M}^\mu {\delta_N}^\nu 
 \left[\sqrt{-{\widehat G}}{\widehat G}_{\mu\nu}{\widetilde f}-
 M_P^{2-D}T_{\mu\nu}\right] \delta^{(2)}(x^i)=0~.
 \label{EoM1}
\end{eqnarray}
where we should keep terms linear in the fluctuations $h_{MN}$, which are
assumed to vanish once we turn off the brane matter source $T_{\mu\nu}$.

{}In fact, the linearized equations of motion are quite simple. The reason
for this simplification is that the background is factorizable, and the 
Gauss-Bonnet terms give contributions only on the brane. Indeed, outside of the
brane the $D$-dimensional space-time locally is Minkowski, and the 
linearized Gauss-Bonnet contributions vanish for flat backgrounds. Thus, it is 
not difficult to show that the linearized equations of motion have the 
following form:
\begin{eqnarray}
 &&R_{MN}-\frac{1}{2}\ G_{MN}R+\nonumber\\
 &&+\displaystyle{
 {1\over2\sqrt{-G}}}
 {\delta_M}^\mu {\delta_N}^\nu 
 \left[\sqrt{-{\widehat G}}{\widetilde f}\left[{\widehat G}_{\mu\nu}+
 4\lambda
 \left({\widehat R}_{\mu\nu}-\frac{1}{2}\ 
 {\widehat G}_{\mu\nu} {\widehat R}\right)\right] -
 M_P^{2-D}T_{\mu\nu}\right] \delta^{(2)}(x^i)=0~.
\label{EoM2}
\end{eqnarray}
Note that these linearized equations are the same as those following from the
action $S_*+S_{\rm{\small int}}$, where
\begin{equation}
 S_*={\widehat M}_P^{D-4}\int_{\Sigma}\ d^{D-2}x\sqrt{-\widehat G}
 \left[{\widehat R}-{\widehat\Lambda}\right]
 +M_P^{D-2}\int\ d^D x\sqrt{-G}~R~,
 \label{model1}
\end{equation}
while
\begin{equation}
 {\widehat M}_P^{D-4}\equiv 2\lambda f~,
\end{equation}
and ${\widehat \Lambda}\equiv 1/2\lambda$. That is, 
at the linearized
level the model (\ref{model}) coincides with the model (\ref{model1}), albeit
they are different beyond the linearized approximation. In particular, the
backgrounds corresponding to the ans{\"a}tz (\ref{ansatz}) 
are the same in both models. 

{}Thus, as far as the linearized level is concerned, the contributions
from the Gauss-Bonnet term are equivalent to having a tree-level 
Einstein-Hilbert term on the brane. As was pointed out in \cite{DGP,DG}, such 
a term drastically modifies the behavior of gravity at short {\em vs.}
long distances. Note that in our case the $(D-2)$-dimensional Planck scale on
the brane is given by ${\widehat M}_P$. Positivity of ${\widehat M}_P^{D-4}$
then requires that $\lambda$ be positive. Finally, note that for 
\begin{equation}
 \exp(-\beta)=1/N~,
\end{equation} 
where $N$ is a positive integer, the wedge is nothing but the
${\bf R}^2/{\bf Z}_N$ orbifold with the origin of the wedge identified as the
orbifold fixed point. 

\subsection{Linearized Equations of Motion}

{}The linearized equations of motion for the fluctuations $h_{MN}$ induced
by the brane matter are given by:
\bea
 &&-\nabla^L\nabla_L h_{MN}+2\nabla^L\nabla_{(M}h_{N)L}
 -\nabla_M\nabla_N h +G^{(0)}_{MN}\left(\nabla^L\nabla_L h
 -\nabla^L\nabla^K h_{LK}\right)=\nonumber\\
 &&{M_P^{2-D}\over
 \sqrt{\widetilde G}}{\widetilde T}_{MN}
 \delta^{(2)}(x^i)~,\label{EoML}
\eea
where the components of the ``effective'' energy-momentum tensor
${\widetilde T}_{MN}$ are given by
\bea
 &&{\widetilde T}_{\mu\nu}=T_{\mu\nu}-
 {\widehat M}_P^{D-4}\biggl[-\partial^\lambda\partial_\lambda H_{\mu\nu}+
 2\partial^\lambda\partial_{(\mu} H_{\nu)\lambda}-\partial_\mu\partial_\nu H
 +\eta_{\mu\nu}\left(\partial^\lambda\partial_\lambda H
 -\partial^\lambda\partial^\sigma H_{\lambda\sigma}
 \right)\biggr]\label{T-mu-nu}~,\\
 &&{\widetilde T}_{\mu i}= f ~h_{\mu i}~,\label{T-mu-n}\\
 &&{\widetilde T}_{ij} = 0~, \label{T-m-n}
\eea
and we have defined $h\equiv h^M{}_M$ and $H\equiv H^\mu{}_\mu$.
These equations of motion are invariant under certain gauge transformations
corresponding to unbroken diffeomorphisms. Since the brane has non-zero 
tension, some of the diffeomorphisms
\bea\label{diff}
 \delta h_{MN}=\nabla_M\xi_N+\nabla_N\xi_M~,
\eea
corresponding to the $D$-dimensional 
reparametrizations
\begin{equation}
 x^M\rightarrow x^M-\xi^M(x)~,
\end{equation}
are actually broken at the origin of the wedge. Thus, it is not difficult
to show that the $(ij)$ components of (\ref{EoML}) are invariant under
the full $D$-dimensional diffeomorphisms (\ref{diff}), while the invariance of
the $(\mu i)$ and the $(\mu\nu)$ components requires that
\bea
 && f\ {\delta^{(2)}(x^k)\over \sqrt{\widetilde G}}~\nabla_i\xi_\mu
 = 0~,
 \nonumber\\
 && f\ \nabla^i \left[\xi_i{\delta^{(2)}(x^k)\over \sqrt{\widetilde G}}
 \right]=0~,
\eea
respectively. Note that these conditions are trivial in the case of a 
tensionless brane (where the space is flat everywhere, including at the
origin). However, for a non-zero tension brane these conditions give
non-trivial restrictions on the gauge parameters at the origin of the wedge
(away from the origin these conditions, once again, are trivial). Thus, we
have:
\begin{eqnarray}\label{cond1}
 &&\partial_i\xi_\mu|_{\rho=0}=0~,\\
 &&\partial^i\left[\xi_i\delta^{(2)}(x^k)\right]=0~.
\end{eqnarray} 
Because of these conditions, some care is needed in gauge fixing in this 
model. In particular, there are subtleties with imposing a gauge such as 
the harmonic gauge (if $f\not=0$). At any rate,
we will solve the above equations of motion without appealing
to such gauge fixing.

{}Since we are looking for solutions to the above linearized equations of 
motion such that $h_{MN}$ vanish for vanishing $T_{\mu\nu}$, it is clear
that the graviphoton components $h_{\mu i}$ must be vanishing everywhere:
\begin{equation}
 h_{\mu i}\equiv 0~.
\end{equation}
Indeed, the graviphotons do not couple to the conserved energy-momentum 
tensor $T_{\mu\nu}$ on the brane. Moreover, the graviscalar components
$\chi_{ij}\equiv h_{ij}$ only couple to the trace of $T_{\mu\nu}$, that is,
$T\equiv T_\mu{}^\mu$. This implies that we have
\begin{equation}
 \chi_{ij}={1\over 2}G^{(0)}_{ij}\chi~,
\end{equation}
where $\chi\equiv (G^{(0)})^{ij}\chi_{ij}$. The equations of motion for
$H_{\mu\nu}$ and $\chi$ then simplify as follows (note that $h=H+\chi$):
\begin{eqnarray}
 &&-\left(\partial^\lambda\partial_\lambda+
 \nabla^i\nabla_i\right) H_{\mu\nu}+2\
 \partial^\lambda\partial_{(\mu}H_{\nu)\lambda}
 -\partial_\mu\partial_\nu H- \partial_\mu\partial_\nu \chi+\nonumber\\
 &&\eta_{\mu\nu}\left(\partial^\lambda\partial_\lambda H+
 \partial^\lambda\partial_\lambda \chi+\nabla^i\nabla_i H+{1\over 2}
 \nabla^i\nabla_i \chi-\partial^\lambda\partial^\sigma H_{\lambda\sigma}
 \right)={M_P^{2-D}\over
 \sqrt{\widetilde G}}{\widetilde T}_{\mu\nu}
 \delta^{(2)}(x^i)~,\label{EoML1}\\
 &&\partial_i\left[\partial^\lambda H_{\mu\lambda}-\partial_\mu H-
 {1\over 2}\partial_\mu
 \chi\right]=0~,\label{EoML2}\\
 &&-\nabla_i\nabla_j H+G_{ij}^{(0)}\left[
 \partial^\lambda\partial_\lambda H+{1\over 2}\partial^\lambda\partial_\lambda
 \chi +\nabla^k\nabla_k H-\partial^\lambda\partial^\sigma H_{\lambda\sigma}
 \right]=0~.
\end{eqnarray}
{}From the last equation it follows that
\begin{eqnarray}
 &&\partial^\lambda\partial_\lambda H+
 {1\over 2}\partial^\lambda\partial_\lambda
 \chi +{1\over 2}
 \nabla^i\nabla_i H-\partial^\lambda\partial^\sigma H_{\lambda\sigma}=0~,\\
 &&\nabla_i\nabla_j H={1\over 2}G^{(0)}_{ij} \nabla^k\nabla_k H~.
\end{eqnarray}
Let us begin discussing this system of equations by studying the
last equation for $H$.

{}This equation can be rewritten as follows:
\begin{equation}
 \partial_i\partial_j H-\partial_i\omega\partial_j H-\partial_j\omega
 \partial_i H=\delta_{ij}\left[{1\over 2}\partial^k\partial_k H-\partial^k
 \omega\partial_k H\right]~.
\end{equation}
Consider axially symmetric solutions: $H=H(x^\mu,\rho)$ (recall that
$\rho^2=x^i x_i$). Then we have:
\begin{equation}
 H^{\prime\prime}+{1\over\rho}(2\nu-1)H^\prime=0~,
\end{equation}
where prime stands for derivative w.r.t. $\rho$. The general solution to this
equation is given by
\begin{equation}
 H(x^\mu,\rho)=B(x^\mu)\left({\rho^2\over a^2}\right)^{1-\nu}+C(x^\mu)~,
\end{equation}
where $B,C$ {\em a priori} are arbitrary functions of $x^\mu$. Note, however,
that since $0<\nu<1$, we must have $B(x^\mu)\equiv 0$. This implies that
$H$ is only a function of $x^\mu$. We can then always gauge it away using
the $(D-2)$-dimensional diffeomorphisms with the gauge parameters 
$\xi_\mu(x^\sigma)$ independent
of $x^i$ (note that such gauge transformations do not affect the graviphoton 
or graviscalar components). 
Thus, we conclude that $H$ can be set to zero everywhere. Note that
this is actually correct even for $\nu=0$, that is, in the case of a
tensionless brane.

{}With $H\equiv 0$ the equations of motion simplify as follows:
\begin{eqnarray}
 &&-\left(\partial^\lambda\partial_\lambda+
 \nabla^i\nabla_i\right) H_{\mu\nu}+2\
 \partial^\lambda\partial_{(\mu}H_{\nu)\lambda}
 - \partial_\mu\partial_\nu \chi+\nonumber\\
 &&\eta_{\mu\nu}\left(
 \partial^\lambda\partial_\lambda \chi+{1\over 2}
 \nabla^i\nabla_i \chi-\partial^\lambda\partial^\sigma H_{\lambda\sigma}
 \right)={M_P^{2-D}\over
 \sqrt{\widetilde G}}{\widetilde T}_{\mu\nu}
 \delta^{(2)}(x^i)~,\label{EoMLL1}\\
 &&\partial_i\left[\partial^\lambda H_{\mu\lambda}-
 {1\over 2}\partial_\mu
 \chi\right]=0~,\label{EoMLL2}\\
 &&
 {1\over 2}\partial^\lambda\partial_\lambda
 \chi -\partial^\lambda\partial^\sigma H_{\lambda\sigma}=0~.
 \label{EoMLL3}
\end{eqnarray}
Also, note that
\bea
 {\widetilde T}_{\mu\nu}&=& T_{\mu\nu}-
 {\widehat M}_P^{D-4}\biggl[-\partial^\lambda\partial_\lambda H_{\mu\nu}+
 2\partial^\lambda\partial_{(\mu} H_{\nu)\lambda}
 -\eta_{\mu\nu}\partial^\lambda\partial^\sigma H_{\lambda\sigma}
 \biggr]\label{T-mu-nu1}~,
\eea
We, therefore, have
\begin{equation}
 {\widetilde T}=T+{{D-4}\over 2}{\widehat M}_P^{D-4}
 \partial^\lambda\partial_\lambda\chi~.
\end{equation}
On the other hand, taking the trace of (\ref{EoMLL1}), we have:
\begin{equation}
 \left[\partial^\lambda\partial_\lambda+\nabla^i\nabla_i\right]\chi=
 {2\over{D-2}}{M_P^{2-D}\over
 \sqrt{\widetilde G}}{\widetilde T}\delta^{(2)}(x^i)~.
\end{equation}
Next, note that (\ref{EoMLL2}) and (\ref{EoMLL3}) imply that
\begin{equation}
 \partial^\lambda H_{\mu\lambda}-
 {1\over 2}\partial_\mu
 \chi={1\over 2} \partial_\mu g~,
\end{equation}
where $g=g(x^\mu)$ is independent of $x^i$, and satisfies the 
$(D-2)$-dimensional Klein-Gordon equation
\begin{equation}\label{f}
 \partial^\lambda\partial_\lambda g=0~.
\end{equation}
It then follows that
\begin{eqnarray}
 -\left(\partial^\lambda\partial_\lambda+
 \nabla^i\nabla_i\right) H_{\mu\nu}+\partial_\mu\partial_\nu g
 ={M_P^{2-D}\over
 \sqrt{\widetilde G}}\left[{\widetilde T}_{\mu\nu}-{1\over{D-2}}\eta_{\mu\nu}
 {\widetilde T}\right]
 \delta^{(2)}(x^i)~,
\end{eqnarray}
where
\begin{equation}
 {\widetilde T}_{\mu\nu}= T_{\mu\nu}-
 {\widehat M}_P^{D-4}\biggl[-\partial^\lambda\partial_\lambda H_{\mu\nu}+
 \partial_\mu\partial_\nu(\chi+g)
 -{1\over 2}\eta_{\mu\nu}\partial^\lambda\partial_\lambda \chi
 \biggr]\label{T-mu-nu2}~.
\end{equation}
We are now ready to solve for $H_{\mu\nu}$ and $\chi$.

{}To do this, let us Fourier transform the coordinates $x^\mu$ on the
brane. Let the corresponding momenta be $p^\mu$, and let $p^2\equiv
p^\mu p_\mu$. Then we have
\begin{eqnarray}
 &&-\left(
 \nabla^i\nabla_i-p^2\right) H_{\mu\nu}-p_\mu p_\nu g
 ={M_P^{2-D}\over
 \sqrt{\widetilde G}}\left[{\widetilde T}_{\mu\nu}(p)-
 {1\over{D-2}}\eta_{\mu\nu}
 {\widetilde T}(p)\right]
 \delta^{(2)}(x^i)~,\\
 &&\left[\nabla^i\nabla_i-p^2\right]\chi=
 {2\over{D-2}}{M_P^{2-D}\over
 \sqrt{\widetilde G}}{\widetilde T}(p)\delta^{(2)}(x^i)~,
\end{eqnarray}
where
\begin{equation}
 {\widetilde T}_{\mu\nu}(p)= T_{\mu\nu}(p)-
 {\widehat M}_P^{D-4}\biggl[p^2 H_{\mu\nu}-p_\mu p_\nu(\chi+g)
 +{p^2\over 2}\eta_{\mu\nu}\chi
 \biggr]\label{T-mu-nu3}~.
\end{equation}
Note that the equation (\ref{f}) now reads
\begin{equation}
 p^2 g=0~,
\end{equation}
so $g\equiv 0$ for $p^2\not =0$. On the other hand, for $p^2=0$ the equation
for $H_{\mu\nu}$ {\em away} from the brane reads:
\begin{equation}
 \nabla^i\nabla_i H_{\mu\nu}=-p_\mu p_\nu g~.
\end{equation}
This gives (for axially symmetric $H_{\mu\nu}$):
\begin{equation}
 H_{\mu\nu}^{\prime\prime}+{1\over \rho} H_{\mu\nu}^\prime=
 -p_\mu p_\nu \left({a^2\over\rho^2}\right)^\nu g~,
\end{equation} 
where $g$ is independent of $\rho$. For non-vanishing $g$ we would then have
\begin{equation}
 H_{\mu\nu}\sim -{p_\mu p_\nu a^2\over 4(1-\nu)^2}
\left({\rho^2\over a^2}\right)^{1-\nu} g
\end{equation}
for large $\rho$. This implies that even for $p^2=0$ we must set $g=0$.

{}Thus, the equations of motion for $H_{\mu\nu}$ and $\chi$ read:
\begin{eqnarray}\label{nonzero}
 &&-\left(
 \nabla^i\nabla_i-p^2\right) H_{\mu\nu}
 ={M_P^{2-D}\over
 \sqrt{\widetilde G}}\left[{\widetilde T}_{\mu\nu}(p)-
 {1\over{D-2}}\eta_{\mu\nu}
 {\widetilde T}(p)\right]
 \delta^{(2)}(x^i)~,\\
 &&\left[\nabla^i\nabla_i-p^2\right]\chi=
 {2\over{D-2}}{M_P^{2-D}\over
 \sqrt{\widetilde G}}{\widetilde T}(p)\delta^{(2)}(x^i)~,
\end{eqnarray}
where
\begin{equation}
 {\widetilde T}_{\mu\nu}(p)= T_{\mu\nu}(p)-
 {\widehat M}_P^{D-4}\biggl[p^2 H_{\mu\nu}-p_\mu p_\nu\chi
 +{p^2\over 2}\eta_{\mu\nu}\chi
 \biggr]\label{T-mu-nu4}~.
\end{equation}
To solve these equations, we must distinguish between the cases where
$p^2=0$ and $p^2\not=0$.

{}Let us start with the $p^2\not=0$ case. Then, due to the fact that
the two-dimensional propagator is logarithmically divergent at the origin,
we have (this is in complete parallel with the discussion in \cite{DG})
\begin{equation}
 H_{\mu\nu}=\chi=0~,~~~\rho\not=0~,
\end{equation}
while on the brane we have
\bea
 &&H_{\mu\nu}(p_\lambda,\rho=0)=\frac{{\widehat M}^{4-D}}
 {p^2}\biggl[T_{\mu\nu}(p)-{1\over
 D-4}\eta_{\mu\nu} T(p)+{2\over D-4}\frac{p_\mu p_\nu}{p^2}T(p) \biggr]
\label{H-mu-nu}~,\\
 &&\chi(p_\lambda,\rho=0)={2\over {D-4}}\frac{{\widehat M}^{4-D}}
 {p^2} T(p)~.\label{chi00}
\eea
Note that, according to these expression, 
the $p^2\not=0$ modes are completely localized on the brane.
We will, however, come back to this point after discussing the
$p^2=0$ case.

{}Thus, in the $p^2=0$ case we have:
\begin{eqnarray}
 &&-
 \partial^i\partial_i H_{\mu\nu}
 =M_P^{2-D}\left[T_{\mu\nu}(p)-
 {1\over{D-2}}\eta_{\mu\nu}
 T(p)+{\widehat M}_P^{D-4} p_\mu p_\nu\chi\right]
 \delta^{(2)}(x^i)~,\\
 &&\partial^i\partial_i\chi=
 {2\over{D-2}}M_P^{2-D}T(p)\delta^{(2)}(x^i)~.
\end{eqnarray}
The solution to the equation for $\chi$ is given by:
\begin{equation}
 \chi(p^2=0,\rho)={M_P^{2-D}\over 2\pi(D-2)} T(p)\ln\left({\rho^2\over b^2}
 \right)~,
\end{equation}
where $b$ is an integration constant. Note that unless $T(p)\equiv 0$, the
solution for $\chi$ is singular at the origin, so that the equation for
$H_{\mu\nu}$ is ill-defined due to the term proportional to
$p_\mu p_\nu\chi$ as the latter blows up at the origin\footnote{Note that
this term does not affect the coupling of the graviton $H_{\mu\nu}$
to the brane matter
as $p^\mu T_{\mu\nu}(p)=0$ for such matter. However, this term can be probed
by bulk matter as $p^\mu T^{\rm{\small bulk}}_{\mu\nu}(p)$ need {\em 
not} be zero.}. Here we would like to emphasize 
that this term
{\em cannot} be removed by a gauge transformation. Since this singularity is
a short-distance singularity, it is expected to be smoothed out by 
ultra-violet effects which we are neglecting here\footnote{For instance, if
the brane has small width instead of being $\delta$-function-like, this 
singularity is absent. Note that, as was pointed out in \cite{DG}, 
in this case complete localization of gravity is not expected to be the case
either. Instead, it is expected that gravity is $(D-2)$-dimensional below
some cross-over distance scale $r_c$ (which depends on the brane width),
while it become $D$-dimensional at distances larger than $r_c$.}. This 
smoothing out can simply be modeled via
\begin{equation}
 \chi(p^2=0,\rho)={M_P^{2-D}\over 2\pi(D-2)} T(p)\ln\left({{\rho^2+\epsilon^2}
 \over b^2}\right)~,
\end{equation}
where $\epsilon$ is a small parameter with the dimension of 
length\footnote{At least in some cases we can expect that $\epsilon\sim
1/\Lambda$, where $\Lambda$ is an ultra-violet cut-off in the theory.}.
This amounts to smoothing out the $\delta$-function source via
\begin{equation}\label{delta}
 \delta^{(2)}(x^i)\rightarrow {1\over\pi}{\epsilon^2\over
 (\rho^2+\epsilon^2)^2}~.
\end{equation}
We then have the following solution for the graviton $H_{\mu\nu}$:
\begin{eqnarray}
 &&H_{\mu\nu}(p^2=0,\rho)=\nonumber\\
 &&=-{M_P^{2-D}\over 4\pi}\left[T_{\mu\nu}(p)-
 {1\over{D-2}}\eta_{\mu\nu}
 T(p)+{\widehat M}_P^{D-4} p_\mu p_\nu\chi(p^2=0,\rho=0)\right]
 \ln\left({{\rho^2+\epsilon^2}
 \over b^{\prime 2}}\right)~,
\end{eqnarray}
where $b^\prime$ is an integration constant.

{}In fact, smoothing out of the aforementioned singularity
also smoothes out a singularity 
in the $p^2\not=0$ case if the brane has non-zero tension. 
Indeed, from (\ref{chi00}) it follows that, if $T(p)\not=0$, $\chi$ is 
non-vanishing on the brane (but it vanishes in the bulk). The corresponding
graviscalar components are given by:
\begin{equation}
 \chi_{ij}={1\over 2}\delta_{ij}\exp(2\omega)\chi~,
\end{equation}  
which are infinite as $\exp(2\omega)$ diverges on the brane if $0<\nu<1$.
However, if we smooth out the $\delta$-function via (\ref{delta}), then
we have
\begin{equation}
 \exp(2\omega)=\left({a^2\over{\rho^2+\epsilon^2}}\right)^\nu~,
\end{equation}
which is now non-singular at $\rho=0$. Note that for a smoothed out brane
the $p^2\not=0$ modes now also penetrate into the bulk as can be seen from
(\ref{nonzero}). However, as was originally pointed out in \cite{DG}, 
for small enough $\epsilon$, only ultra-light modes penetrate into the bulk 
efficiently (that is, with a substantial wave-function in the bulk).

\subsection{The Tensionless Brane Case}

{}The conclusions of the previous subsection are applicable in the case of 
a tensionless brane. In this case we can arrive at the same conclusions in a
somewhat simpler way. Thus, consider the codimension-2 Dvali-Gabadadze model:
\begin{equation}
 S={\widehat M}^{D-4}_P\int_\Sigma d^{D-2}x \sqrt{{\widehat G}}{\widehat R}+
 {M}^{D-2}_P\int d^{D}x \sqrt{{G}}{R}~.
\end{equation}
The background in this model is flat: $G^{(0)}_{MN}=\eta_{MN}$. The linearized
equations of motion for the fluctuations $h_{MN}$ are given by:
\bea
 &&\biggl[-\partial^L\partial_L h_{MN}+2\partial^L\partial_{(M}h_{N)L}
 -\partial_M\partial_N h +\nonumber\\
 &&\hskip3cm \eta_{MN}\left(\partial^L\partial_L h
 -\partial^L\partial^K h_{LK}\right)\biggr]=M_P^{2-D}{\widetilde T}_{MN}
 \delta^{(2)}(x^i)~,\label{EoMLT}
\eea
where the components of the ``effective'' energy-momentum tensor
${\widetilde T}_{MN}$ are given by
\bea
 {\widetilde T}_{\mu\nu}= T_{\mu\nu}-
 {\widehat M}_P^{D-4}\biggl[-\partial^\lambda\partial_\lambda H_{\mu\nu}+
 2\partial^\lambda\partial_{(\mu} H_{\nu)\lambda}-\partial_\mu\partial_\nu H
 +\eta_{\mu\nu}\left(\partial^\lambda\partial_\lambda H
 -\partial^\lambda\partial^\sigma H_{\lambda\sigma}
 \right)\biggr]\label{T-mu-nuT}~,
\eea
while ${\widetilde T}_{\mu i}$ and ${\widetilde T}_{ij}$ are zero.

{}Note that, since the brane is tensionless, the full $D$-dimensional
diffeomorphisms are intact:
\begin{equation}
 \delta h_{MN}=\partial_M\xi_N+\partial_N\xi_M~.
\end{equation}
We can, therefore, use the harmonic gauge:
\begin{equation}\label{harm}
 \partial^M h_{MN}={1\over 2}\partial_N h~.
\end{equation}
This gives:
\bea
 &&-\partial^L\partial_L h_{MN}+{1\over 2}
 \eta_{MN}\partial^L\partial_L h=M_P^{2-D}{\widetilde T}_{MN}
 \delta^{(2)}(x^i)~.\label{EoMLT1}
\eea
It then follows that
\begin{equation}
 -\partial^L\partial_L h_{MN}=M_P^{2-D}\left[{\widetilde T}_{MN}-{1\over{D-2}}
 \eta_{MN} {\widetilde T}\right]
 \delta^{(2)}(x^i)~.
\end{equation}
Once again, the graviphoton components vanish ($h_{\mu i}=0$), while for the
graviscalar components we have 
\begin{equation}\label{chichi}
 \chi_{ij}={1\over 2}\delta_{ij}\chi~.
\end{equation} 
This together with the harmonic gauge (\ref{harm}) implies that
\begin{eqnarray}
 &&\partial^\mu H_{\mu\nu}={1\over 2}\partial_\nu H+ {1\over 2}\partial_\nu
 \chi~,\\
 &&\partial_j H=0~.
\end{eqnarray}
The latter allows to set $H\equiv 0$, and
\begin{eqnarray}
 &&\partial^\mu H_{\mu\nu}={1\over 2}\partial_\nu
 \chi~.
\end{eqnarray}
In particular, we have
\bea
 {\widetilde T}_{\mu\nu}&=& T_{\mu\nu}-
 {\widehat M}_P^{D-4}\biggl[-\partial^\lambda\partial_\lambda H_{\mu\nu}+
 \partial_\mu\partial_\nu\chi
 -{1\over 2}\eta_{\mu\nu}\partial^\lambda\partial_\lambda \chi
 \biggr]\label{T-mu-nuT1}~.
\eea
The equations of motion simplify as follows:
\begin{eqnarray}
 &&-\left(
 \partial^i\partial_i-p^2\right) H_{\mu\nu}
 ={M_P^{2-D}}\left[{\widetilde T}_{\mu\nu}(p)-
 {1\over{D-2}}\eta_{\mu\nu}
 {\widetilde T}(p)\right]
 \delta^{(2)}(x^i)~,\\
 &&\left[\partial^i\partial_i-p^2\right]\chi=
 {2\over{D-2}}{M_P^{2-D}}{\widetilde T}(p)\delta^{(2)}(x^i)~,
\end{eqnarray}
where
\bea
 {\widetilde T}_{\mu\nu}(p)&=& T_{\mu\nu}(p)-
 {\widehat M}_P^{D-4}\biggl[p^2 H_{\mu\nu}-
 p_\mu p_\nu\chi+
 {p^2\over 2}\eta_{\mu\nu} \chi
 \biggr]\label{T-mu-nuT2}~,
\eea
and we have Fourier transformed the coordinates $x^\mu$. Note that 
these equations are
precisely the same as at the end of the previous subsection for the case
where the transverse space is flat. Note that, just as in the case of a
non-zero tension brane, for the $p^2=0$ modes the 
$p_\mu p_\nu\chi$ term in ${\widetilde T}_{\mu\nu}$ is still singular on the
brane. This singularity is removed once we smooth out the $\delta$-function 
as in (\ref{delta}). On the other hand, for a strictly $\delta$-function-like
brane this singularity in the {\em linearized} theory would lead to 
inconsistencies somewhat similar to those discussed in \cite{COSM}, which, 
in particular, could be probed by bulk matter (see footnote 5). 
Note, however, that 
the presence of this term indicates that the
linearized theory might be breaking down, which would imply that a more
complete 
non-perturbative analysis (which is outside of the scope of this paper), say,
along the lines suggested in 
\cite{DDGV} might be required here\footnote{This singularity might be 
analogous to that arising in a linearized theory of a massive graviton
as discussed in \cite{DDGV}.}. If, however, this inconsistency 
persists non-perturbatively in the case of a (both tensionless as well
as non-zero tension) strictly $\delta$-function-like brane, 
it appears that we would have to appeal to smoothing out via ultra-violet
physics. 
It would be interesting to understand this
point better\footnote{Here we note that consistent infinite-volume
brane world scenarios with non-conformal brane matter were
recently discussed in the string theory context in
\cite{radu,CIKL}. These are generalizations of
their conformal counterparts of \cite{KS,LNV,BKV,orient}. Their orientifold
generalizations can also be discussed, but some caution is needed due to
the issues discussed in \cite{KaSh,KST}. The brane Einstein-Hilbert term 
in the string theory context was discussed in 
\cite{alberto,Lowe,Kiritsis,CIKL}.}.

\acknowledgments

{}We would like to thank Gia Dvali and Gregory Gabadadze for
valuable discussions.
This work was supported in part by the National Science Foundation and
an Alfred P. Sloan Fellowship. Parts of this work were completed during
Z.K.'s visit at New York University.
Z.K. would also like to thank Albert and Ribena Yu for financial support.


\begin{references}

\bibitem{early}
V. Rubakov and M. Shaposhnikov, Phys. Lett. {\bf B125} (1983) 136.

\bibitem{BK}
A. Barnaveli and O. Kancheli, Sov. J. Nucl. Phys. {\bf 52} (1990) 576.

\bibitem{polchi} J. Polchinski, Phys. Rev. Lett. {\bf 75} (1995) 4724.

\bibitem{witt} P. Ho{\u r}ava and E. Witten, Nucl. Phys. {\bf B460} (1996)
506; Nucl. Phys. {\bf B475} (1996) 94;\\
E. Witten, Nucl. Phys. {\bf B471} (1996) 135.

\bibitem{lyk} I. Antoniadis, Phys. Lett. {\bf B246} (1990) 377;\\
J. Lykken, Phys. Rev. {\bf D54} (1996) 3693.

\bibitem{shif} G. Dvali and M. Shifman, Nucl. Phys. {\bf B504} (1997) 127;
Phys. Lett. {\bf B396} (1997) 64.

\bibitem{TeV} N. Arkani-Hamed, S. Dimopoulos and G. Dvali,
Phys. Lett. {\bf B429} (1998) 263; Phys. Rev. {\bf D59} (1999) 086004.

\bibitem{dienes} K.R. Dienes, E. Dudas and T. Gherghetta, Phys. Lett.
{\bf B436} (1998) 55; Nucl. Phys. {\bf B537} (1999) 47; hep-ph/9807522;\\
Z. Kakushadze, Nucl. Phys. {\bf B548} (1999) 205; 
Nucl. Phys. {\bf 551} (1999) 549;
Nucl. Phys.
{\bf B552} (1999) 3;\\
Z. Kakushadze and T.R. Taylor, Nucl. Phys. {\bf B562} (1999) 78.

\bibitem{3gen} Z. Kakushadze, Phys. Lett. {\bf B434} (1998) 269;
Nucl. Phys. {\bf B535} (1998) 311; Phys. Rev. {\bf D58} (1998) 101901.

\bibitem{anto} I. Antoniadis, N. Arkani-Hamed, S. Dimopoulos and G. Dvali,
Phys. Lett. {\bf B436} (1998) 257.

\bibitem{ST} G. Shiu and S.-H.H. Tye, Phys. Rev. {\bf D58} (1998) 106007.

\bibitem{BW} Z. Kakushadze and S.-H.H. Tye, Nucl. Phys. {\bf B548} (1999) 180;
Phys. Rev. {\bf D58} (1998) 126001.

\bibitem{Gog} M. Gogberashvili, hep-ph/9812296; Europhys. Lett. {\bf 49}
(2000) 396.

\bibitem{RS} L. Randall and R. Sundrum, Phys. Rev. Lett. {\bf 83} (1999)
3370; Phys. Rev. Lett. {\bf 83} (1999) 4690.

\bibitem{kogan} I.I. Kogan, S. Mouslopoulos, A. Papazoglou, G.G. Ross and 
J. Santiago, Nucl. Phys. {\bf B584} (2000) 313.
 
\bibitem{DGP} G. Dvali, G. Gabadadze and M. Porrati, Phys. Lett. {\bf
B485} (2000) 208.

\bibitem{DG} G. Dvali and G. Gabadadze, Phys. Rev. {\bf D63} (2001) 065007.

\bibitem{alberto} A. Iglesias and Z. Kakushadze, hep-th/0011111;
hep-th/0012049.

\bibitem{GRS} R. Gregory, V.A. Rubakov and S.M. Sibiryakov,
Phys. Rev. Lett. {\bf 84} (2000) 5928.

\bibitem{CEH} C. Csaki, J. Erlich and T.J. Hollowood, Phys. Rev. Lett. {\bf
84} (2000) 5932.

\bibitem{DGP0} G. Dvali, G. Gabadadze and M. Porrati, Phys. Lett. {\bf B484}
(2000) 112; Phys. Lett. {\bf B484} (2000) 129.

\bibitem{witten} E. Witten, hep-ph/0002297.

\bibitem{DVALI} G. Dvali, hep-th/0004057.

\bibitem{zura} Z. Kakushadze, Phys. Lett. {\bf B488} (2000) 402;
Phys. Lett. {\bf B489} (2000) 207; Phys. Lett. {\bf B491} (2000) 317;
Mod. Phys. Lett. {\bf A15} (2000) 1879.

\bibitem{CP} A. Chodos and E. Poppitz, Phys. Lett. {\bf B471} (1999) 119.

\bibitem{GS} T. Gherghetta and M. Shaposhnikov, Phys. Rev. Lett.
{\bf 85} (2000) 240.

\bibitem{olzu2} O. Corradini and Z. Kakushadze, Phys. Lett. {\bf B506} (2001)
167.

\bibitem{COSM} Z. Kakushadze, Nucl. Phys. {\bf B589} (2000) 75;
Phys. Lett. {\bf B497} (2001) 125;\\
O. Corradini and Z. Kakushadze, Phys. Lett. {\bf B494} (2000) 302;\\
Z. Kakushadze and P. Langfelder, Mod. Phys. Lett. {\bf A15} (2000) 2265.

\bibitem{DDGV} C. Deffayet, G. Dvali, G. Gabadadze and A. Vainshtein,
hep-th/0106001.

\bibitem{radu} Z. Kakushadze and R. Roiban, JHEP {\bf 0103} (2001) 043.

\bibitem{CIKL} O. Corradini, A. Iglesias, Z. Kakushadze and P. Langfelder,
hep-th/0107167.

\bibitem{KS} S. Kachru and E. Silverstein, Phys. Rev. Lett. {\bf 80}
(1998) 4855.

\bibitem{LNV} A. Lawrence, N. Nekrasov and C. Vafa,
Nucl. Phys. {\bf B533} (1998) 199.

\bibitem{BKV}
M. Bershadsky, Z. Kakushadze and C. Vafa, Nucl. Phys. {\bf B523}
(1998) 59.

\bibitem{orient}
Z. Kakushadze, Nucl. Phys. {\bf B529} (1998) 157;  Phys. Rev. {\bf D58}
(1998) 106003; Phys. Rev. {\bf D59} (1999) 045007; Nucl. Phys. {\bf B544}
(1999) 265.

\bibitem{KaSh} Z. Kakushadze, Nucl. Phys. {\bf B512} (1998) 221;\\
Z. Kakushadze and G. Shiu, Phys. Rev. {\bf D56} (1997) 3686;
Nucl. Phys. {\bf B520} (1998) 75.

\bibitem{KST}Z. Kakushadze, G. Shiu and S.-H.H. Tye,
Nucl. Phys. {\bf B533} (1998) 25;\\
Z. Kakushadze,
Phys. Lett. {\bf B455} (1999) 120; Int. J. Mod. Phys. {\bf A15} (2000) 3461;
Phys. Lett. {\bf B459} (1999) 497; Int. J. Mod. Phys. {\bf A15} (2000)
3113.

\bibitem{Lowe} S. Corley, D. Lowe and S. Ramgoolam, JHEP {\bf 0107} (2001) 030.

\bibitem{Kiritsis} E. Kiritsis, N. Tetradis and T. Tomaras, hep-th/0106050.

\end{references}
\end{document}